\documentclass[11pt]{article} 
\usepackage[paperheight=9.44in,paperwidth=6.69in,left=1.85cm,right=1.8cm,bottom=2.5cm,top=3cm,twoside]{geometry}

\usepackage{subfig} 

\usepackage{authblk} 
\usepackage{fancyhdr} 
\usepackage{lastpage} 
\usepackage{textcomp} 
\usepackage{amsmath} 
\usepackage{amssymb} 
\usepackage{amsthm} 
\usepackage{graphicx} 
\usepackage{multirow} 
\usepackage{hyperref} 
\usepackage{longtable} 
\usepackage{float} 
\usepackage{listings} 
\newcommand{\linia}{\noindent\rule{\linewidth}{0.5mm}\hrulefill} 

\usepackage{times}
\usepackage{titlesec}

\titleformat*{\section}{\large\bfseries}
\titleformat*{\subsection}{\normalsize\bfseries}

\fancypagestyle{firststyle}
{
	\fancyhf{}
	\fancyfoot{}
	\fancyhead{}
	\fancyhead[CE]{\upshape Zeszyty Naukowe WWSI, No 21, Vol. 13, 2018, pp. \thepage-\pageref{LastPage} \newline DOI: 10.26348/znwwsi.21.74}
	\fancyhead[CO]{\upshape Zeszyty Naukowe WWSI, No 21, Vol. 13, 2018, pp. \thepage-\pageref{LastPage} \newline DOI: 10.26348/znwwsi.21.74}
	\fancyfoot[L,R]{\footnotesize\bfseries Manuscript received August 18, 2019}
	\fancyfoot[LE,RO]{\thepage}
	}

\title{\large \bfseries Detection of a Source Code Plagiarism \\in a Student Programming Competition} 
\author{\normalsize Zenon Gniazdowski\thanks{E-mail: zgniazdowski@wwsi.edu.pl} }
\author{\normalsize Maciej Boniecki}
\affil{\normalsize Warsaw School of Computer Science}

\date{\vspace{-5ex}}
\providecommand{\keywords}[1]{\textbf{\textit{Keywords ---}} #1}

\begin{document}

\maketitle 
\thispagestyle{firststyle} 

\linia
\begin{abstract}\label{abstract}
	\noindent The article presents a system for testing the independence of solutions to algorithmic problems sent by students as part of the student programming competition. First, the context was discussed, as well as the need to organize programming competitions resulting from this context. Then, an algorithm was proposed to study the mutual similarity of source codes of programs sent as part of a programming competition. Since, after implementation, the algorithm was used in practice, examples of its application for detecting the plagiarism of source codes of solutions in two programming competitions conducted as part of classes on Algorithms and Numerical Methods were also presented. Finally, the effectiveness of the solutions used in the work was discussed.
\end{abstract}

\section{Introduction}
One of the elements of information technology studies is teaching the basics of programming. By programming, the authors of the article understand a certain sequence of activities that lead to the creation of a program that is used to solve some practical problem. The sequence described here will be a simplified model of the programming process and probably within it there will be some nuances that we will not consider in this article. However, within this sequence, in a intuitive way, certain steps will be identified that lead to the creation of a computer program:
\begin{itemize}
	\item First the problem to be solved should be identified;
	\item Then an algorithm should be found that will solve the problem;
	\item In turn, the algorithm should be written in the form of a program in a certain programming language;
	\item Finally, the program should be started and tested.
\end{itemize}
While teaching programming, there are some problems that interfere with this process. These problems should be prevented. An important way to prevent problems is continuous programming, similar to sports training. For this purpose, you can use the dedicated programming competition on the portal offering competitions.

Students should fairly participate in the competition. However, one cannot forget that participation in the program competition may be dishonest. This may manifest itself, for example, in plagiarizing the solutions of other participants. Plagiarism is not uncommon, but this is not a local case either. Whole scientific conferences are devoted to plagiarism \cite{Glendinning2017}.
We can look at plagiarism from different points of view.  On the one hand, it is talked about in the context of school systems \cite{Cosma2006} \cite{Culwin2001}, on the other hand, it is talked about in the context of cultural differences \cite{Bamford2005}\cite{Doss2016}\cite{Gow2014}\cite{Shei2005}.

This article considers possible problems that occur in teaching programming and proposes ways to counteract these problems, both from the point of view of the student's goals and from the point of view of the teacher's goals.
In particular, a simple method was proposed for testing the level of plagiarism of source code from other participants of the programming competition. The effectiveness of the proposed method was also analyzed.
\section{Preliminaries}
The preliminaries will present some problems that occur in the programming teaching process. These can be workshop problems of a technical nature, but also mental problems. Programming competitions that can be an antidote to these problems will also be presented. The issues presented here will provide a context for the presentation of further activities undertaken as part of programming teaching classes.
\subsection{Problems encountered while learning programming}
Students starting the process of learning the basics of programming have very diverse experience. It can be assumed that this experience falls within a range limited by two extremes. On the one hand, they are people who have considerable knowledge and skills in programming. On the other hand, there are people who have no experience in programming, and as a result of lack of education, they have too little knowledge of the basics of mathematics, and in particular too little knowledge of mathematical logic.

Teaching students who do not have basic mathematical and logical knowledge is a challenge. In order to meet this challenge, one should more accurately diagnose and list potential difficulties in the programming teaching process. If we assume that these students can fill gaps in mathematical and logical knowledge as part of mathematics classes, there are still two groups of difficulties to overcome. The first group of difficulties are workshop difficulties. The second group is mental difficulties.
\subsubsection{Workshop difficulties}
Programming presupposes knowledge of some programming language as well as knowledge of algorithms. It can usually be assumed that at the very beginning of the course the student knows a certain programming language. Also, the need to learn a new programming language should not be a problem as long as the student sees analogies between different languages: between their dictionaries and also between their grammars. Perhaps a slightly bigger problem is the change in the programming paradigm with the change of language. However, it seems that this problem also disappears with the experience gained.

In turn, programming requires knowledge of algorithms and the ability to consciously use them. Here the matter is more complex. On the one hand, there are difficult problems for which effective algorithms are not known. For example, the problem of integer factorization is such a difficult problem.
On the other hand, many algorithms have been discovered and described. Their description can be found in textbooks, in professional magazines or on the Internet. In this case, it is enough to read, understand and consciously encode them.
\subsubsection{Mental problems}
At the initial stage of programming learning, it can be assumed that neither the problems solved are too complex nor the language structures used are very complicated. Despite this, in many cases students learning programming can still see some inability. This is followed by an internal conviction that this inability cannot be overcome. The student is convinced that he can not solve the problem, because the problem is too complex and exceeds its capabilities. He can't solve it now and will never be able to solve it. So he has the right to think he's of little value.

Is there a way out of this circle of impossibility? An analogy with sport comes to mind immediately. Nobody gets record results at the beginning of their sports path. The results may appear later if they are preceded by appropriate training. By training, the technique improves systematically, but the body's capabilities also increase. Training starts with the simplest elements. As the technique improves and the body's capabilities increase, workouts are expanded with new elements. Over time, comparing his previous results with what he is achieving now, the athlete notices that his confidence increases, and therefore his self-esteem increases. Initial weakness was overcome.

The same can be done when learning programming. At the beginning of the road you should start by solving numerous, but the simplest programming tasks. Over time, more and more complex problems will obviously be solved, but more and more complex programming structures will be used to solve them. With successive problems solved, successive degrees of know-how will be achieved. Over time, the initial circle of inability will be weakened or even completely eliminated. In this way, existing mental barriers will be broken and self-esteem will return.
\subsection{Programming competitions}
In learning programming, web portals that offer the ability to solve programming tasks are useful. There are many such portals. Examples are: http://solve.edu.pl https://www.spoj.com/. After registering, you can solve the competition tasks offered by the portal. Each task contains a description of the problem to be solved, sample test data and the correct program result for that data. After logging in, a task is selected, the solution of which will be a working program written in one of the available programming languages. Using the available interface, the source code of the written program is sent to the system. The system will check the correctness of the submitted solution. If the source code you submitted contains syntax errors, the system will inform you about it. The solution can be sent again after correcting the errors. When there are no syntax errors, the program is compiled and then repeatedly run for different sets of test data. The program results are compared with the reference results. If the result of the program is correct for all test sets, then a message about accepting the task appears. Otherwise, error messages will appear. It can be a message saying that the program gave the wrong answer, or that its operation exceeded the allowable time limit, i.e. the program did not finish within the specified time interval, or that another error occurred while the program was running.
With such messages, it is still possible to go back to the source code and improve it \cite{Boniecki2009}.
\section{Dedicated programming competitions in algorithm teaching}
The first author of this article at Computer Engineering Studies conducts Algorithm classes for students of the third semester, and also conducts classes in Numerical Methods for students of the fifth semester.
It should be emphasized that determining the final grade for each participant of these classes is a big challenge.
In this context, two facts are important:
\begin{itemize}
	\item Portal https://www.spoj.com/ offers programming competitions.
	\item On this portal, the second author of this article has administrative privileges, while the first author of this article has judicial privileges.
\end{itemize}
It was decided to use these facts in didactics. For the students taking part in the abovementioned classes, two competitions were organized in the winter semester of the 2018/2019 academic year. The algorithmic competition was available at https://www.spoj.com/WWSIASD/, while the numerical competition was available at https://www.spoj.com/WWSIMN/.
The competition tasks included solving problems for which effective algorithms are known and described. It was only necessary to know these algorithms, to understand them and to code them themselves. Solutions for individual tasks could be sent to the checking system from the moment these tasks were visible in the competition participant's panel. Solving time was set individually for each task. Usually this time ranged from seven to fourteen days. At the end of the semester, based on the competition ranking, a list of people who passed the exam was proposed. These people also received the final grade from the classes.

As the experiment was considered successful, similar competitions were organized in the winter semester of the 2019/2020 academic year.
\subsection{The problem of reliability of final grades}\label{Absurdy}
Because the competition tasks concerned well-known problems, the participants of the competition could search for appropriate algorithms in textbooks, in professional magazines or on the Internet. Cooperation between competition participants was not excluded. This cooperation could include, for example, help in understanding the algorithm. The only practical limitation in solving tasks is that their solutions are not thoughtlessly copied using the "copy" and "paste" shortcuts.

Although cooperation is permissible, one should be aware that it may also carry some dangers related to the integrity of solutions. The student group participating in the class consists of many people. Different people will represent different attitudes towards the group's standards. The group will include people whose attitude will oppose breaking the norms. However, it cannot be ruled out that those who break the standards will also be found there. Since social relations should be based on trust, a lecturer conducting group classes should assume the good will of all group members. This assumption has its rational justification:
\begin{itemize}
	\item For the average person, deliberate violation of established norms leads to internal anxiety.
	\item Paying tuition fees for learning and rejecting learning opportunities is nonsensical from an economic point of view.
\end{itemize}
On the other hand, the assumption of good will of class participants does not absolve the lecturer from checking the integrity of the work they send. In particular, it should be examined whether the solutions sent are not plagiarism of the work of other participants of the competition. This is not an easy issue, if only because in some cases an identical or almost identical solution will not indicate plagiarism, but rather that the algorithm cannot be written significantly differently. An example of such a situation would be coding of some well-known algorithm, e.g. Euclid's algorithm for finding the largest common divisor of two positive integers. This simple algorithm has the form:
\begin{equation}\label{Eq01}
	gcd(a,b)=\begin{cases}
		a,              & \text{if $b = 0$}. \\
		gcd(b,a\mod b), & \text{otherwise}.
	\end{cases}
\end{equation}
In the case of an algorithm with such a simple structure, it can be assumed with a high degree of certainty that in a given programming language its implementations will not differ significantly. Unfortunately, hence the conclusion that testing the independence of solutions cannot be limited to a simple analysis of the similarity of the source code of the programs. The assessment of the solution should also take into account the context in which the similar fragments of the source code occur.
\section{Detection of source code plagiarism in programming competitions}
In response to the problem of plagiarism, there are methods and tools that try to counteract this problem \cite{Ali2011}\cite{Hage2010}\cite{Heres2017}\cite{Lukashenko2007}\cite{Martins2014}. This article also proposes some way of identifying plagiarism. A program has been developed that allows the analysis of mutual similarity between the source codes of the programs that were sent for evaluation.
The proposed solution is relatively simple, and thus does not take into account many particularly perfidious threats. However, the authors of the article assumed that in a situation where the deadline for sending source codes of solved programming problems is approaching, and therefore in a situation of permanent time pressure, the participants of the competition will not be able to use more sophisticated methods of plagiarism.

\subsection{A measure of similarity between source codes}\label{Similarity}
First of all, the dissimilarity of the source codes of the programs submitted for verification in the competition was analyzed. For this purpose, the algorithm of finding the edit distance between two ASCII strings was used \cite{Dasgupta2008}.
At the input of the algorithm (called the Levenshtein algorithm) were given two properly prepared strings of characters. Each string represented one of the source codes. The result of the program was the Levenshtein distance, informing about the difference between the two strings, measured by the smallest number of simple operations needed to transform one string into another.

Preliminary preparation of string should prevent potential errors in the analysis of their (dis)similarities:
\begin{itemize}
	\item The programs compared were mostly written in C++ language. In turn, any program written in this language requires the application of specific preprocessor directives. On the other hand, without affecting the correct operation of the program, the preprocessor directives could be added to confuse the algorithm that compares the source codes of the two programs. Because the preprocessor directives do not affect the similarity level of the two source codes, therefore, all preprocessor directives, i.e. all strings beginning with the {"\#"} character, were omitted in the analysis.
	\item Also, the comments do not affect the correct operation of the program. On the other hand, comments can be formed freely, if only to deceive the program examining the similarity of two source codes. Therefore, in the analyzed source codes all comments were omitted, both those preceded by the pair of characters "//",  as well as comments between the separators "/*" and "*/".
	\item Two identical source codes can be differently formatted, i.e. they can differ by any number of whitespace. Because the excess of whitespace does not affect the correctness of the program code, and may confuse the algorithm comparing the two source codes, so before finding the edit distance of these codes, all whitespace was removed from them.
\end{itemize}
After the initial preparation of the source codes, their mutual edit distance $d_L$ was calculated. At the known edit distance of two strings, if the length of the strings being compared is not known, it cannot be clearly stated how (dis)similar the two strings are. With short strings, the similarity will be large, and with long strings, the similarity will be small.
It seems that instead of the edit distance, a relative measure of similarity in the $\left[ 0,1\right] $ range would be better.
Levenshtein similarity \cite{Deza2009} could be such a measure:
\begin{equation}\label{Eq02}
	S_L=1-\frac{d_L}{max \left( m,n\right) }
\end{equation}
The following designations have been adopted in the above formula:
\begin{itemize}
	\item $S_L$ - Levenshtein similarity,
	\item $d_L$ - Levenshtein edit distance,
	\item $m, n$ - lengths of compared strings.
\end{itemize}
Since Levenshtein similarity is a number in the range of $\left[ 0,1\right] $, they can also be expressed as a percentage.

For the purposes of assessing the similarity of the two source codes, the above measure of similarity was used. This measure was estimated for a given task, for all pairs of its solutions that were accepted by the checking system.
In this way, for a given task, for $k$ analyzed source codes from $k$ different students, a square symmetrical similarity matrix with the size $k \times k$ was obtained. The values in the matrix were in the range of $\left[ 0,1\right] $. On its diagonal were unit values.
\subsection{Assessment of plagiarism propensity}\label{Propensity}
A given row of Levenshtein similarity matrix contains levels of similarity of the solution sent by a given student with solutions sent by all other students.
The values in the rows of this similarity matrix are numbers from the range $\left[ 0,1\right] $.
From the analysis of formula (\ref{Eq02}) it can be concluded that for Levenshtein's similarity of two strings to reach zero value, their edit distance must be equal to the greater of the two lengths of the compared strings. Intuition suggests that in the case of program source codes, zero similarity values are unlikely.

The unit values on the diagonal of the matrix describe the similarity of the source codes sent for assessment to themselves. Non-diagonal values in a given row of the similarity matrix characterize the similarity of the source code sent by a given competitor to the source codes of other competitors. At least one of the non-diagonal values takes the maximum value. It can be assumed that this maximum value, denoted as $P$, characterizes the student's propensity for plagiarism.
If the $P$ value is small, it means that the solution sent is significantly different from the other solutions. In the opposite case, the participant could not or did not want to send an independent solution to the task in the competition. In the first case, the propensity for plagiarism is imperceptible. In the second case, it is significant.

Finding the values of plagiarism propensity for all students participating in the competition, using descriptive statistics tools, it is possible to find features of the plagiarism propensity distribution, specific for the whole group of competition participants.
\subsection{The relation defined at a given level of similarity}\label{Relacja}
For further analysis, the maximum threshold of acceptable similarity was arbitrarily assumed. If the similarity value exceeded this threshold, it was assumed that the two source codes were too similar to recognize the independence of the solution. In this way, the case of plagiarism was identified. Otherwise, the plagiarism problem was considered to be non-existent.
It can be assumed that those solutions whose similarity is greater than a given threshold are in mutual relation with each other. Otherwise, the relation does not occur. In this way, the relation matrix \cite{Ross1992} can be formed from the similarity matrix. If any codes are in relation to other source codes, they are the result of plagiarism. Those source codes that are not related to any other were not plagiarized.
The relation matrix is a binary symmetrical matrix containing unit values on the diagonal. This means that this relation is at least reflexive and symmetrical, so it is a relation of similarity \cite{Peters2012}. If additionally it is a transitive relation, it is also an equivalence relation.
Using the relation matrix, you can draw a graph of it that shows the existing relations between the analyzed source codes. The graph will consist of connected components. Each connected component represents the similarity class, while the similarity class represents the groups of competitors whose cooperation consisted in copying the finished source code. If the similarity class (connected component of the graph) contains only one node, then the participant of the competition corresponding to this node of the graph, submitted for assessment a solution of a task that is not similar to other solutions. This solution is not plagiarism.

A simple analysis of the similarity graph is not enough to evaluate the solutions submitted. Graph nodes should be clustered. Based on the obtained adjacency matrix of the relation graph, adjacency lists can be created. By using these adjacency lists, the DFS algorithm can identify connected components of a graph \cite{Dasgupta2008}. Each connected component represents a list of similar solutions, i.e. those solutions that plagiarize each other.

\subsection{The problem of the final grade for the competition}\label{Grade}
The proposals presented above give the opportunity to find interdependencies between the solutions submitted so far. These results can be used to give a final grade for participation in the competition. The easiest way to get the final ranking is to analyze the number of tasks solved. If the competition participant receives one point for each correctly solved task, then the ranking list of competition participants may be formed after the competition. By setting the boundary, students can be divided into those who passed the classes and those who need to improve them. On the basis of the ranking, those who passed the classes can be assigned final grades for participation in the classes.

The problem arises when the phenomenon of plagiarism is identified in the group. In this case, students plagiarizing should be disqualified or at least should lower their position in the final ranking. Here are two solutions. The first solution links the score for the task, with the hypothetical contribution of the participants of a given plagiarizing cluster to solving the task. If more than one student has been identified in the cluster, assuming an even contribution of all cluster participants to the solution of this task, each student can be awarded only the part of the task score that is proportional to its contribution to the solution.
For a cluster consisting of $n$ students, each of the students belonging to a given cluster may be awarded with a prize of $1/n$ point for their contribution to the solution of a given task:
\begin{equation}\label{Eq03}
	G=\frac{1}{n}.
\end{equation}
In the second solution, the student's grade for a given task sent in the competition can be made dependent on his propensity to plagiarize $P$ (see subsection \ref{Propensity}).
There can be at least two ways to evaluate the task submitted in the competition:
\begin{itemize}
	\item Student can be assessed binary. With the propensity $P$ not less than the assumed threshold $\varepsilon$, the $G$ grade obtained for the task will be zero points, otherwise one point will be awarded:
	      \begin{equation}\label{Eq04}
		      G=\begin{cases}
			      1, & \text{if $P<\varepsilon$}. \\
			      0, & \text{otherwise}.
		      \end{cases}
	      \end{equation}
	      This method is analogous to the method used to form the relation matrix in subsection \ref{Relacja}.
	\item If the propensity $P$ is not less than the assumed threshold $\varepsilon$, the student can be assigned a $G$ grade which is the difference between the grade possible in the absence of plagiarism ($S=1$) and the similarity:
	      \begin{equation}\label{Eq05}
		      G=\begin{cases}
			      1,   & \text{if $P < \varepsilon$}. \\
			      1-p, & \text{otherwise}.
		      \end{cases}
	      \end{equation}
	      In the absence of plagiarism, the similarity value $S$ will practically never reach zero. Therefore, for sufficiently low values of similarity $S$ ($S<\varepsilon$), the author of the solution is awarded one point per task.
\end{itemize}
\subsection{Algorithm for detecting source code plagiarism}\label{Algorytm}
The analysis presented above allows to describe in detail the algorithm for testing plagiarism of source codes of programs written as part of a programming competition. The algorithm consists of the following steps:
\begin{enumerate}
	\item For a given competition task, read all the source codes of accepted solutions and transform them into a string of characters that can be processed by the Levenshtein algorithm (see subsection \ref{Similarity}). When loading each of the source codes, the following elements should be eliminated from the code:
	      \begin{itemize}
		      \item Preprocessor instructions;
		      \item Comments;
		      \item Whitespace (spaces, tabs, end of line characters).
	      \end{itemize}
	\item For all pairs of transformed $K_i$ and $K_j$ source codes such that $j>i$, using the Levenshtein algorithm, find the edit distance $d_L$ \cite{Dasgupta2008}. Complete the edit distance matrix: $D_{ij} = D_{ji} = d_L$.
	\item Based on the edit distance matrix $D$, using the formula (\ref{Eq02}) estimate the similarity matrix $S$;
	\item On the basis of the $S$ similarity matrix, for participants of the competition who submitted the correct solution, find the plagiarism propensity vector $P$ (see subsection \ref{Propensity}). Subsequent elements of the plagiarism propensity vector $P$ will contain the maximum non-diagonal values of subsequent rows of the $S$ matrix:
	      \begin{equation}\label{Eq06}
		      P_i=\max_{j,j\ne i}S_{ij}.
	      \end{equation}
	\item Establish an $\varepsilon$ plagiarism limit. Using the $\varepsilon$ value, form the relation matrix $R$ from the similarity matrix $S$:
	      \begin{equation}\label{Eq07}
		      R_{ij}=\begin{cases}
			      1, & \text{if $S_{ij}\ge \varepsilon$}. \\
			      0, & \text{otherwise}.
		      \end{cases}
	      \end{equation}
	\item Using the $R$ graph adjacency matrix (\ref{Eq07}), form the graph's adjacency lists.
	\item Using the adjacency lists, use the DFS algorithm \cite{Dasgupta2008} to find connected components of a relation graph. Nodes belonging to a given connected component will represent the source codes of mutually plagiarized programs.
	\item Based on the number of nodes belonging to subsequent connected components (Formula (\ref{Eq03})) or on the basis of the propensity to plagiarism vector $P$ (Formulas (\ref{Eq04}) or (\ref{Eq05})), offer subsequent competitors a score for a given task sent for assessment.
\end{enumerate}
\section{Examples}
\begin{figure}
	\centering
	\includegraphics[width=13cm]{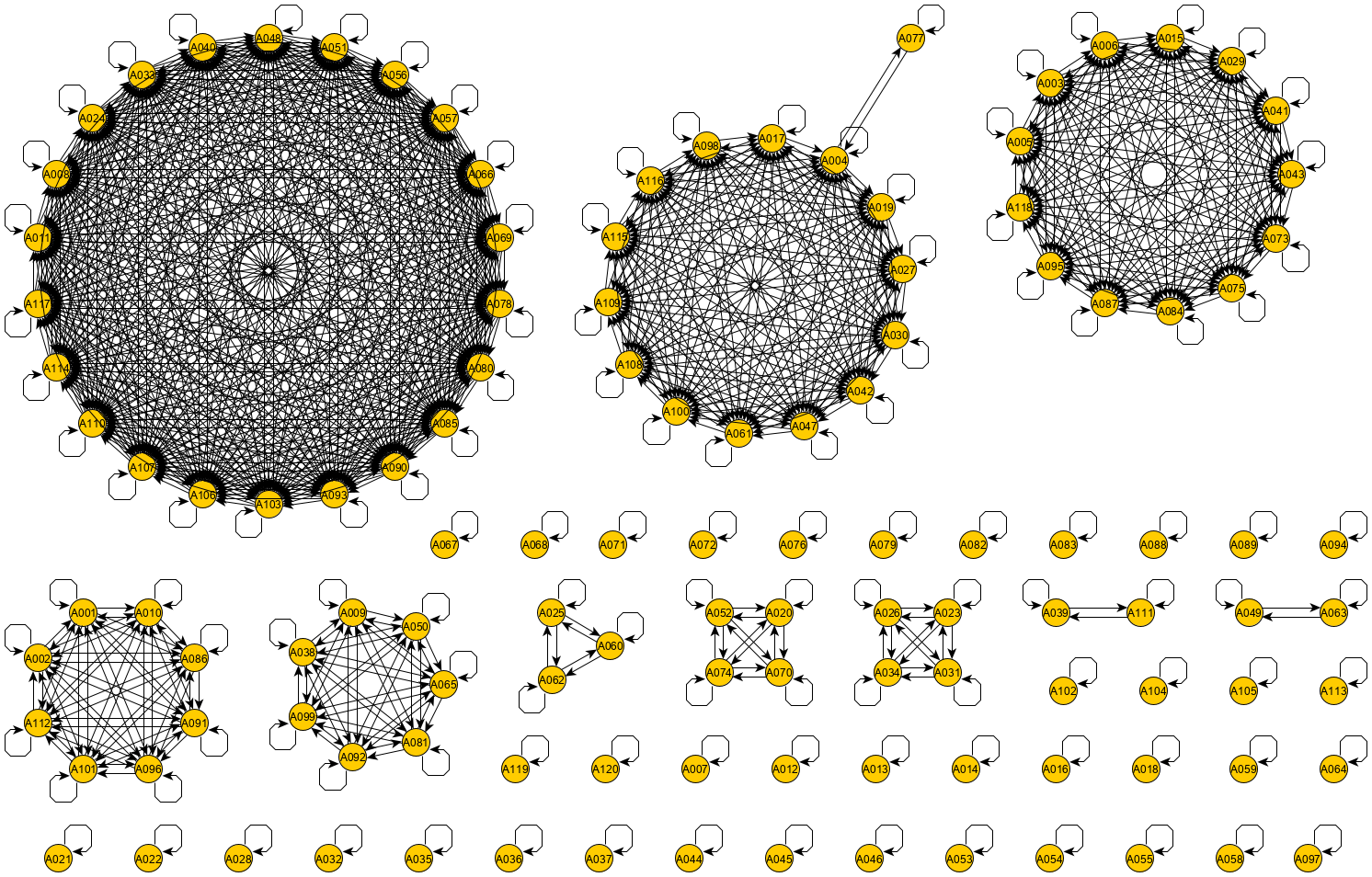}
	\caption{Similarity relation graph for the solutions submitted. Sample algorithmic task from the competition in the 2018/2019 academic year}\label{Graf2018ASD}
\end{figure}

\begin{figure}
	\centering
	\includegraphics[width=11cm]{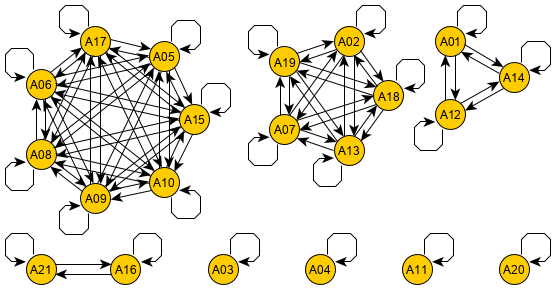}
	\caption{Similarity relation graph for the solutions submitted. Sample numerical problem from the competition in the 2018/2019 academic year}\label{GrafMN2018}
\end{figure}

\begin{figure}
	\centering
	\includegraphics[width=10cm]{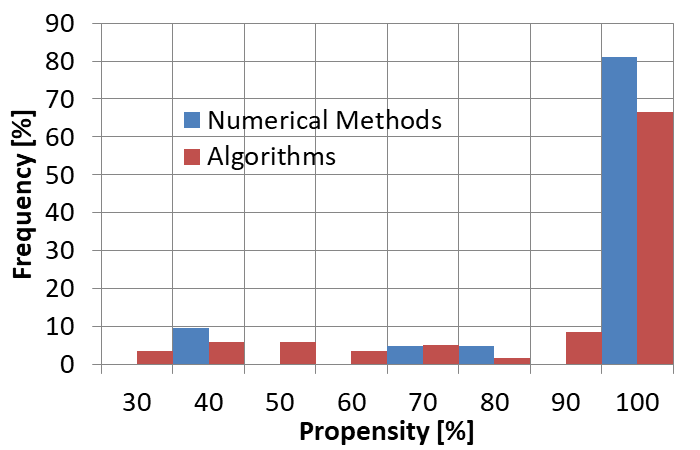}
	\caption{Comparison of histograms for plagiarism propensity during classes on Algorithms and Numerical Methods. Competition in the 2018/2019 academic year.}\label{HistASDiMNprop2018}
\end{figure}

Examination competitions from classes in Algorithms and Numerical Methods were initiated in the winter semester of the 2018/2019 academic year. While reading the source codes of some solutions, it was noticed that some of them are very similar to each other. This persuaded the lecturer to prepare an appropriate program for testing the independence of solutions, described in subsection \ref{Algorytm}. Because the script was not yet ready, which was to extract the source codes from the competition system, therefore the files were manually extracted. Source codes have been extracted for one task from the algorithmic competition (120 correct solutions) and for one task from the numerical competition (21 correct solutions).
For both cases, an analysis of plagiarism was performed. It was found that there are many source codes whose mutual similarity is not less than $90\%$. The graph of similarity relation for source codes from Algorithms is presented in Figure \ref{Graf2018ASD}, and the graph of similarity relation for source codes from Numerical Methods is presented in Figure \ref{GrafMN2018}. The histograms of plagiarism propensity observed during classes from Algorithms and Numerical Methods are also compared. Figure \ref{HistASDiMNprop2018} shows the results of this comparison.

The results obtained were not optimistic. A lot of plagiarisms were sent for evaluation in competitions, despite the fact that before the start of the competitions the teacher encouraged students to cooperate in recognizing algorithms, but also urged them to be independent in coding them. The students understood the first part of his appeal. They forgot the second part. Instead of collaboration, there was cheating \cite{Mason2018}.

Because the analysis was made for one algorithmic problem and for one numerical problem, therefore the lecturer's activities could not be as radical as in the case of analyzing all solutions. However, all students who found themselves in the circle of plagiarizing clusters suffered the consequences. Their final grades for participation in classes were reduced to the level of the lowest positive grades, allowing them to pass the classes.

Before the start of competitions in the 2019/2020 academic year, the lecturer conducting the classes talked to students about plagiarism. This time he emphasized the absurdities of plagiarism presented in subsection \ref{Absurdy}. Because the software analyzing plagiarism has already been enriched with a script that from the competition system allows the automatic extraction of source codes of submitted solutions, therefore in the current edition of the competitions, an analysis of their honesty may be carried out for all submitted solutions.

At the beginning of December, that is at the time of finalizing this article, the competitions reached their halfway point. An analysis of the level of plagiarism was performed for the source codes of the competition programs sent at the moment. In the following subsections of this article, two representative examples of this analysis will be presented.
\subsection{The task in the algorithmic competition in the 2019/20 academic year \\- example of plagiarism analysis}
Up to seventeen programming tasks can be solved in an algorithmic competition. Because the competition has not ended yet, only the nine tasks that were completed were analyzed. As an example, a task that has been solved by 166 students will be presented.

\begin{figure}
	\centering
	\includegraphics[width=10cm]{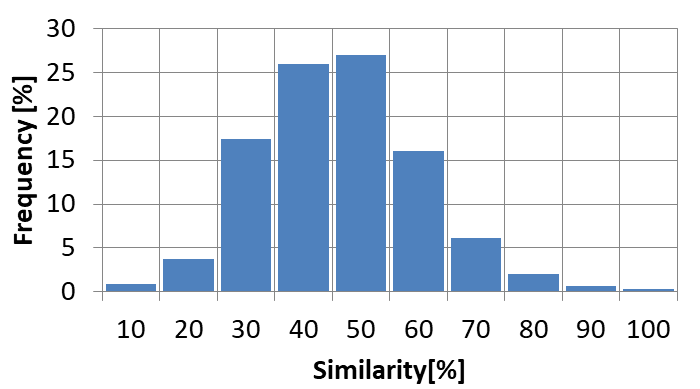}
	\caption{Histogram of similarity distribution of different source codes. The algorithmic task from the competition in the academic year 2019/2020}\label{HistFullASD}
\end{figure}

All source codes of correct solutions submitted have been subjected to anti-plagiarism analysis. According to the procedure described above, the Levenshtein distance and Levenshtein similarity were calculated for each pair of codes (see subsection \ref{Similarity}). There were 13695 of all compared different pairs containing different elements in a pair. For all pairs, both for distance and similarity, selected statistics were calculated (Table \ref{tabASDfullStat}). Figure \ref{HistFullASD} presents the histogram for the code similarity distribution.

\begin{table}
	\centering
	\caption{Statistics calculated for both Levenshtein distance and Levenshtein similarity. The algorithmic task from the competition in the academic year 2019/2020}\label{tabASDfullStat}
	\fontsize{9.5}{13.5}\selectfont{
		\begin{tabular}{c||c|c} \hline
			Statistics         & Levenshtein distance & Levenshtein similarity \\ \hline \hline
			Average            & $137.5$              & $41.34\%$              \\ \hline
			Median             & $95$                 & $40.70\%$              \\ \hline
			Standard deviation & $145.6$              & $14.16\%$              \\ \hline
			Range              & $866$                & $92.42\%$              \\ \hline
			Minimum            & $0$                  & $7.58\%$               \\ \hline
			Maximum            & $866$                & $100\%$                \\ \hline
		\end{tabular}}
\end{table}

\begin{figure}
	\centering
	\includegraphics[width=13cm]{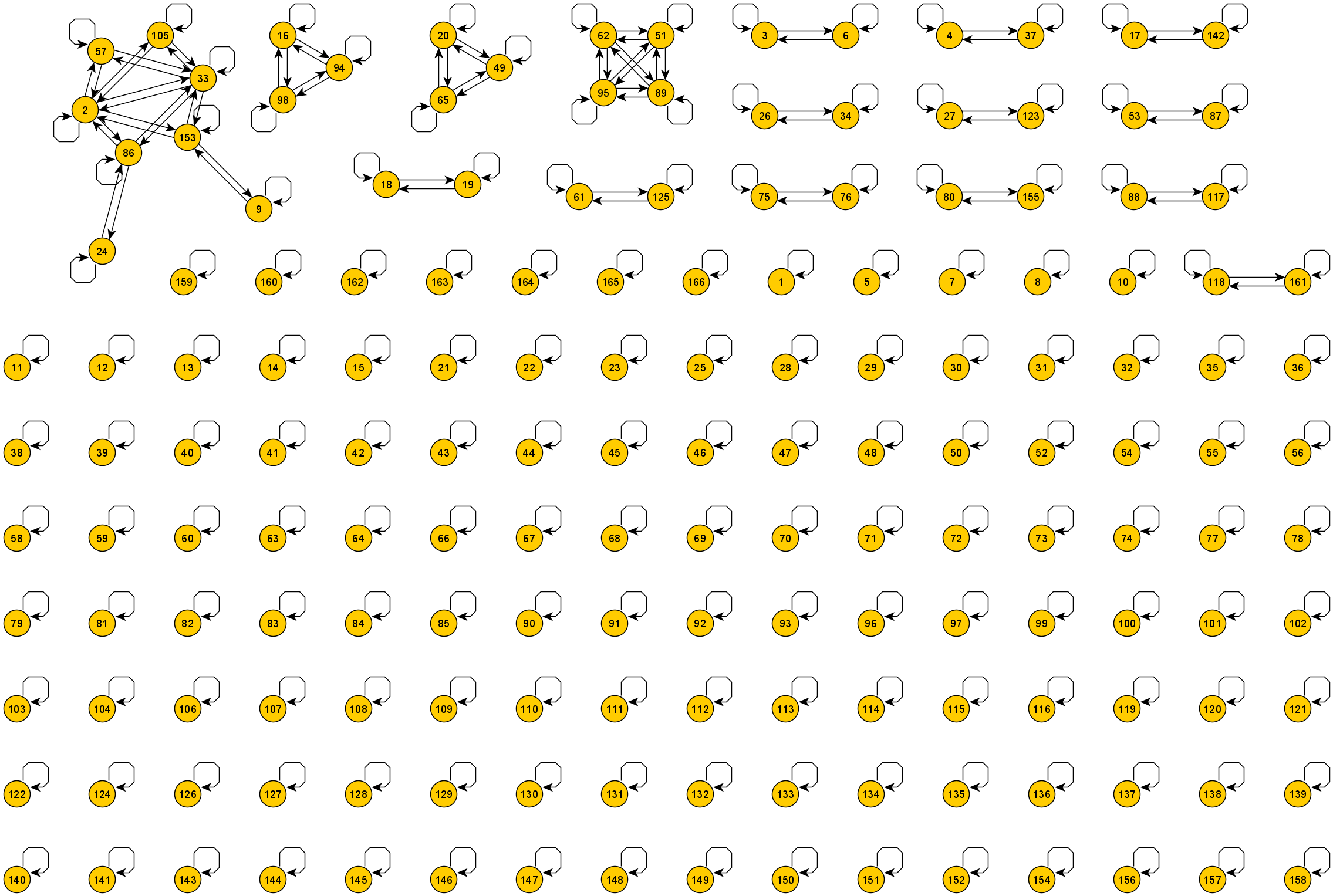}
	\caption{Similarity relation graph for the solutions submitted. The algorithmic task from the competition in the academic year 2019/2020}\label{Graf2019ASD}
\end{figure}

\begin{table}
	\centering
	\caption{Distribution of nodes between clusters. The algorithmic task from the competition in the academic year 2019/2020}\label{tabASDdistOfNodes}
	\fontsize{9.5}{13.5}\selectfont{
		\begin{tabular}{c||c} \hline
			The types of clusters        & The content of the clusters                                           \\ \hline \hline
			$8$ nodes in the cluster     & $\{2,9,24,33,57,86,105,153\}$                                         \\  \hline
			$4$ nodes in the cluster     & $\{51,62,89,95\}$                                                     \\  \hline
			$2$ clusters with $3$ nodes  & $\{16,94,98\}$, $\{20,49,65\}$                                        \\  \hline
			$12$ clusters with $2$ nodes & $\{3,6\}$, $\{4,37\}$, $\{17,142\}$, $\{18,19\}$,                     \\
			                             & $\{26,34\}$, $\{27,123\}$, $\{53,87\}$, $\{61,125\}$,                 \\
			                             & $\{75,76\}$, $\{80,155\}$, $\{88,117\}$, $\{118,161\}$                \\  \hline
			$124$ single-node clusters   & $1$, $5$, $7$, $8$, $10$, $11$, $12$, $13$, $14$, $15$,               \\
			                             & $21$, $22$, $23$, $25$, $28$, $29$, $30$, $31$, $32$, $35$,           \\
			                             & $36$, $38$, $39$, $40$, $41$, $42$, $43$, $44$, $45$, $46$, $47$,     \\
			                             & $48$, $50$, $52$, $54$, $55$, $56$, $58$, $59$, $60$, $63$, $64$,     \\
			                             & $66$, $67$, $68$, $69$, $70$, $71$, $72$, $73$, $74$, $77$, $78$,     \\
			                             & $79$, $81$, $82$, $83$, $84$, $85$, $90$, $91$,  $92$, $93$, $96$,    \\
			                             & $97$, $99$, $100$, $101$, $102$, $103$, $104$, $106$, $107$, $108$,   \\
			                             & $109$, $110$, $111$, $112$, $113$, $114$, $115$, $116$, $119$, $120$, \\
			                             & $121$, $122$, $124$, $126$, $127$, $128$, $129$, $130$, $131$, $132$, \\
			                             & $133$, $134$, $135$, $136$, $137$, $138$, $139$, $140$, $141$, $143$, \\
			                             & $144$, $145$, $146$, $147$, $148$, $149$, $150$, $151$, $152$, $154$, \\
			                             & $156$, $157$, $158$, $159$, $160$, $162$, $163$, $164$, $165$, $166$  \\  \hline
		\end{tabular}}
\end{table}

\begin{table}
	\centering
	\caption{Statistics calculated for the level of plagiarism propensity. The algorithmic task from the competition in the academic year 2019/2020}\label{tabASDPropStat}
	\fontsize{9.5}{13.5}\selectfont{
		\begin{tabular}{c||c} \hline
			Statistics         & Propensity level \\ \hline \hline
			Average            & $75.8\%$         \\ \hline
			Median             & $76.4\%$         \\ \hline
			Standard deviation & $17.3\%$         \\ \hline
			Range              & $71.3\%$         \\ \hline
			Minimum            & $28.7\%$         \\ \hline
			Maximum            & $100\%$          \\ \hline
		\end{tabular}}
\end{table}

\begin{figure}
	\centering
	\includegraphics[width=10cm]{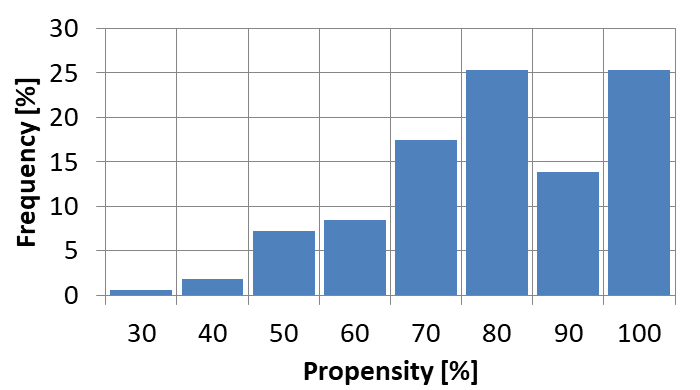}
	\caption{Histogram of plagiarism propensity distribution. The algorithmic task from the competition in the academic year 2019/2020}\label{HistPropASD}
\end{figure}

Finding a good border between plagiarism and non-plagiarism is a major challenge. Since there is no explicit criterion for what level of similarity can be considered evidence of plagiarism, a similarity threshold of $90\%$ was applied. Certainly, below the threshold may be solutions that were plagiarism. However, it seems that a lesser mistake is to underestimate plagiarism than to overestimate. The acceptable level adopted in this study is of common sense. A relation matrix was formed for the adopted similarity threshold. The relation is reflexive and symmetrical, but it is not a transitive relation (Figure \ref{Graf2019ASD}). That means it's a relation of similarity \cite{Peters2012}. The graph's connected components represent similarity classes of this relation.
Each node in the graph represents one valid solution presented by one student. Each edge in the graph represents plagiarism between two different solutions.
Each connected component of the graph represents a group of students who plagiarized the solution together.

Simple graph analysis is not enough to evaluate solutions. Graph nodes should be clustered. Therefore, based on the obtained adjacency matrix, adjacency lists were formed. These adjacency lists were used in the DFS algorithm \cite{Dasgupta2008}  to find graph's connected components. In this way, lists of similar solutions were obtained, i.e. plagiarized solutions were identified. Table \ref{tabASDdistOfNodes} shows the identified distribution of nodes between clusters.

Based on the similarity matrix obtained for the solutions of a given task, the plagiarism propensity vector was estimated. Statistics were also estimated for this vector (Table \ref{tabASDPropStat}). Figure \ref{HistPropASD} also shows a histogram of plagiarism propensity distribution.

After the analysis, it would be possible to proceed to the final assessment for the solved task. However, given the fact that the algorithmic competition has not yet been closed, it has been decided that the final score for the tasks will be awarded only after the end of the competition in order to more reliably take into account all circumstances that may affect the final grades.

\subsection{Problem in the numerical competition in the 2019/20 academic year \\- an example of plagiarism analysis}
Up to thirteen programming problems can be solved in a competition run as part of Numerical Method classes. At the time of finalizing this article at the beginning of December 2019, the competition reached its halfway point and is still ongoing. Therefore, only eight tasks were analyzed, which were completed. As an example, a problem will be presented that has been solved by $6$0 students.

\begin{table}
	\centering
	\caption{Statistics calculated for both Levenshtein distance and Levenshtein similarity. Numerical problem from the competition in the academic year 2019/2020}\label{tabMNfullStat}
	\fontsize{9.5}{13.5}\selectfont{
		\begin{tabular}{c||c|c} \hline
			Statistics         & Levenshtein distance & Levenshtein similarity \\ \hline \hline
			Average            & $753.64$             & $44.42\%$              \\ \hline
			Median             & $912$                & $29.44\%$              \\ \hline
			Standard deviation & $445.16$             & $31.42\%$              \\ \hline
			Range              & $1452$               & $85.33\%$              \\ \hline
			Minimum            & $0$                  & $14.67\%$              \\ \hline
			Maximum            & $1452$               & $100.00\%$             \\ \hline
		\end{tabular}}
\end{table}

\begin{figure}
	\centering
	\includegraphics[width=10cm]{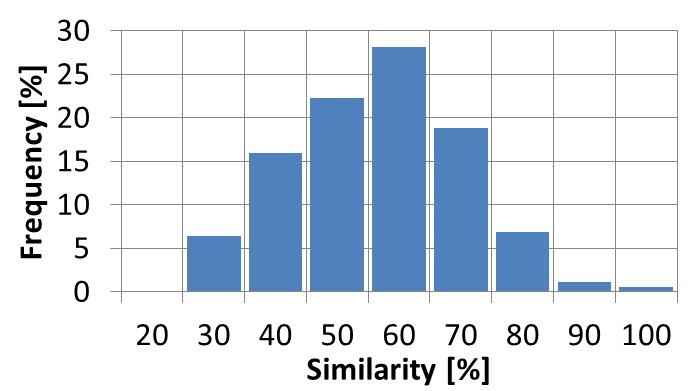}
	\caption{Histogram of the similarity distribution of different source codes. Numerical problem from the competition in the academic year 2019/2020}\label{HistFullMN}
\end{figure}

\begin{figure}
	\centering
	\includegraphics[width=11cm]{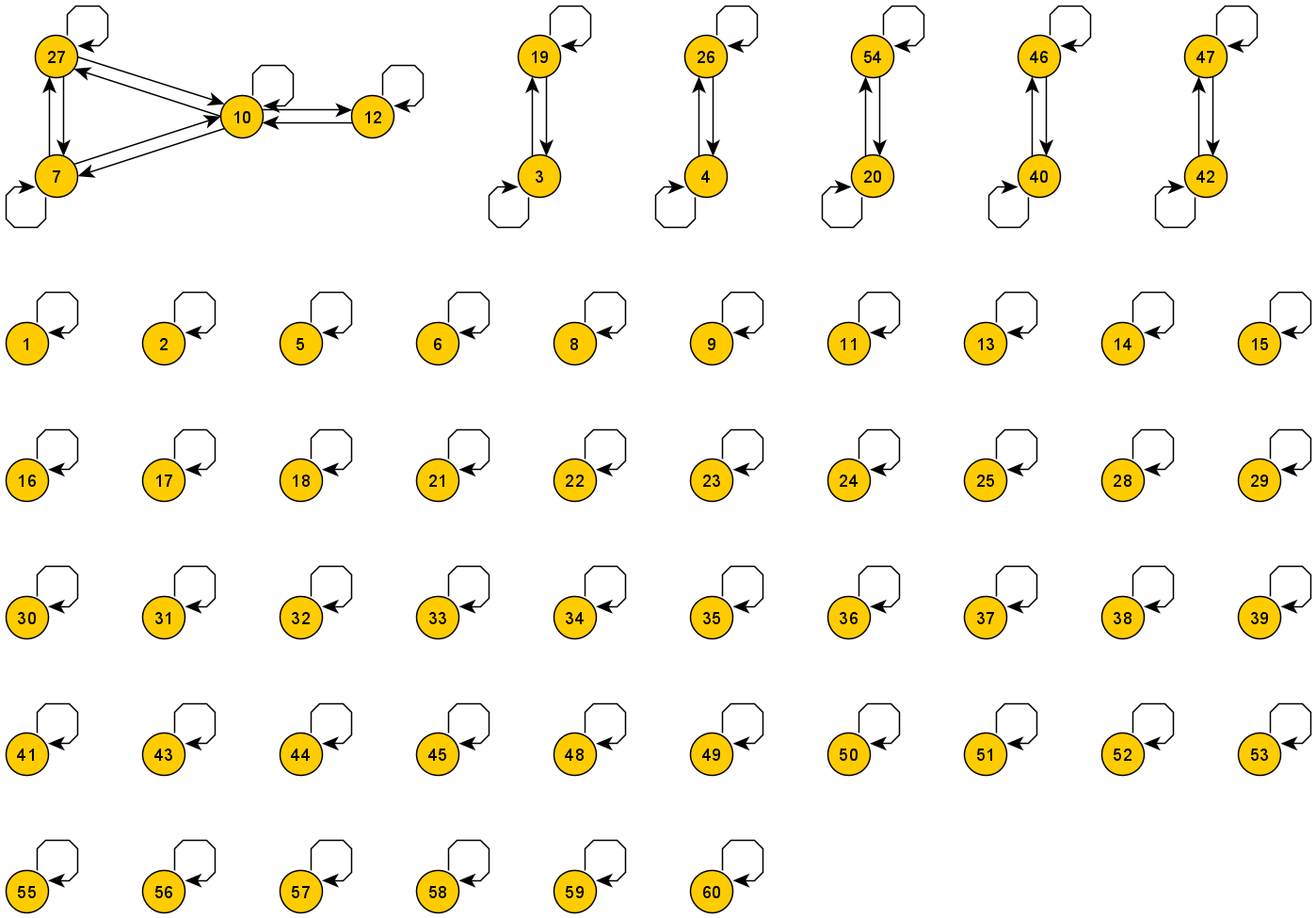}
	\caption{Similarity relation graph for the solutions submitted. Numerical problem from the competition in the academic year 2019/2020}\label{Graf2019MN}
\end{figure}

\begin{figure}
	\centering
	\includegraphics[width=10cm]{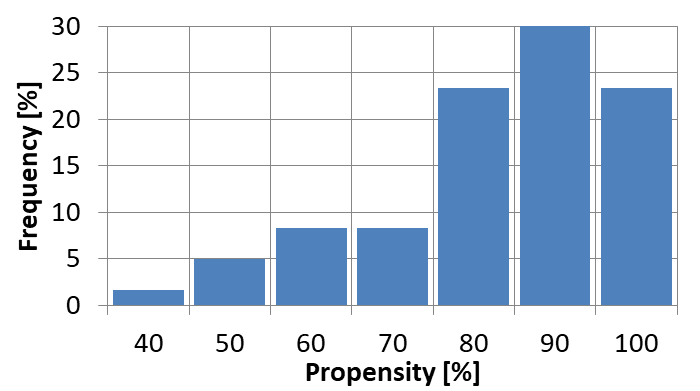}
	\caption{Histogram of plagiarism propensity distribution. The algorithmic task from the competition in the academic year 2019/2020}\label{HistPropMN}
\end{figure}

\begin{table}
	\centering
	\caption{Distribution of nodes between clusters. Numerical problem from the competition in the academic year 2019/2020}\label{tabMNdistOfNodes}
	\fontsize{9.5}{13.5}\selectfont{
		\begin{tabular}{c||c} \hline
			The types of clusters       & The content of the clusters                                             \\ \hline \hline
			$4$ nodes in the cluster    & \{7, 10, 12, 27\}                                                       \\ \hline
			$5$ clusters with $2$ nodes & \{3, 19\}, \{4, 26\}, \{20, 54\} , \{40, 46\}, \{42, 47\}               \\ \hline
			$46$ single-node clusters   & $1$, $2$, $5$, $6$, $8$, $9$, $11$, $13$, $14$, $15$, $16$, $17$, $18$, \\
			                            & $21$, $22$, $23$, $24$, $25$, $28$, $29$, $30$, $31$, $32$, $33$,       \\
			                            & $34$, $35$, $36$, $37$, $38$, $39$, $41$, $43$, $44$, $45$, $48$,       \\
			                            & $49$, $50$, $51$, $52$, $53$, $55$, $56$, $57$, $58$, $59$, $60$        \\ \hline
		\end{tabular}}
\end{table}

\begin{table}
	\centering
	\caption{Statistics calculated for the level of plagiarism propensity. The numerical problem from the competition in the academic year 2019/2020}\label{tabMNPropStat}
	\fontsize{9.5}{13.5}\selectfont{
		\begin{tabular}{c||c} \hline
			Statistics         & Propensity level \\ \hline \hline
			Average            & $78.39\%$        \\ \hline
			Median             & $81.66\%$        \\ \hline
			Standard deviation & $15.27\%$        \\ \hline
			Range              & $62.10\%$        \\ \hline
			Minimum            & $37.90\%$        \\ \hline
			Maximum            & $100\%$          \\ \hline
		\end{tabular}}
\end{table}

In accordance with the procedure described above, all source codes of correct solutions that were sent for evaluation were subjected to anti-plagiarism analysis. Levenshtein distance and Levenshtein similarity were calculated for each pair of codes. There were $1770$ of all compared different pairs containing different elements in pair. Both for Levenshtein distance and for Levenshtein similarity selected statistics were calculated (Table \ref{tabMNfullStat}). Figure \ref{HistFullMN} presents the histogram for the similarity distribution of the source codes sent for evaluation.

In this case, a safe plagiarism level of $90\%$ was also adopted, for which a similarity relation matrix was obtained. Figure \ref{Graf2019MN} presents a graph of this relation containing similarity classes. As in the previous competition, the node in the graph represents one correct solution sent. Each edge in the graph represents the similarity between two different solutions.

Based on the obtained adjacency matrix, adjacency lists were formed, which were used in the DFS algorithm \cite{Dasgupta2008}  to find connected components of the graph. In this way, lists of similar solutions were obtained, i.e. clusters containing plagiarized solutions were identified. Table \ref{tabMNdistOfNodes} shows the distribution of nodes between clusters.

As before, for the algorithmic task, also for the numerical problem, for propensity to plagiarism, statistics were estimated (Table \ref{tabMNPropStat}), and the histogram was presented (Figure \ref{HistPropMN}).

Since the plagiarism analysis of the task was performed, it would also be possible to proceed to final assessment for the completed task. As the numerical competition was also not closed, it was also decided that the final score for the tasks would be awarded only after the end of the competition. As with the algorithmic competition, the goal of this decision is to be more reliable in considering all circumstances that may affect the final results of the competition.

\section{Conclusions}
This article presents a system for testing the independence of source codes sent by students as part of the student programming competition. In particular, the context and the resulting need to organize algorithmic competitions were discussed.
It has been noticed that there are some difficulties in programming teaching. On the one hand, these are workshop difficulties resulting from a lack of knowledge or experience. On the other hand, there are mental difficulties, manifested in the lack of faith in one's own abilities. Both types of difficulty can be overcome by systematically programming.
Programming competitions may encourage systematic programming. It can be seen that the situation in programming teaching is analogous to that in sport, when systematic training, on the one hand, improves efficiency and technique, and on the other increases self-esteem.

Participation in competitions has its good points. However, it may be accompanied by pathological phenomena such as plagiarism. It has been noted that such situations can be countered by the awareness that certain actions may be unfair. If such awareness is lacking, then there is still awareness of the lack of economic sense for such behavior. Plagiarism is an easy solution, but it also takes away the opportunity to grow. Therefore, the anti-plagiarism attitude can be supported by the awareness that money is spent on tuition fees, while at the same time rejecting the possibility of personal development.

Because the level of maturity of students varies, so also their sensitivity and level of self-awareness vary. Hence the certainty that plagiarism will continue to exist. Therefore, it should also be fought with technical tools. This article proposes a systematic approach to this problem. For this purpose, an algorithm has been proposed which uses the Levenshtein edit distance and the concept of Levenshtein similarity.
The algorithm allows to assess the mutual similarity of two different source codes, as well as to assess the propensity for plagiarism.
In addition, a similarity relation can be defined based on the similarity matrix that allows clustering of source codes into groups of mutually similar codes.
The proposed algorithm has been implemented and tested on two examples. It was found that with its help it is possible to identify cases of plagiarism.

By the way, one should notice the overall positive impact of the work being carried out. Since the problem of source code plagiarism was studied over a certain period of time, the effectiveness of anti-plagiarism activities could be seen. The first, preliminary studies were conducted in the 2018/2019 academic year. The second research was conducted during the academic year 2019/2020. It was noted that from year to year the integrity of the source codes sent significantly improved. In most cases, the problem of plagiarism in the 2019/2010 academic year was significantly smaller than in the competitions of the 2018/2019 academic year. Of course, there is no certainty here whether it was influenced by the teaching activities undertaken by the lecturer or the fear of competition participants from detecting their dishonesty. However, due to the final positive effect, it doesn't matter much. Perhaps both activities had an impact on this.

It remains to answer the question whether further activities can be undertaken in the area described in the work. In particular, it seems that in the future it can also be examined whether spectral clustering methods using a similarity matrix can be used for clustering.

\section*{Acknowledgments}
The first author is the main author of the article. He was a moderator in programming competitions. He served as a judge and prepared most of the competition tasks. He also wrote the plagiarism test software presented in the article. Using this software, he studied levels of plagiarism. He also wrote the article.

The second author was the administrator of the competition system. He initiated competitions as well as all competition tasks. He also wrote a script for the extraction of source codes sent to the competition, given to the input of the software used to analyze the level of plagiarism.

\bibliography{zdalne}\label{bibliography}

\begin{thebibliography}{10}
\providecommand{\url}[1]{#1}
\csname url@samestyle\endcsname
\providecommand{\newblock}{\relax}
\providecommand{\bibinfo}[2]{#2}
\providecommand{\BIBentrySTDinterwordspacing}{\spaceskip=0pt\relax}
\providecommand{\BIBentryALTinterwordstretchfactor}{4}
\providecommand{\BIBentryALTinterwordspacing}{\spaceskip=\fontdimen2\font plus
\BIBentryALTinterwordstretchfactor\fontdimen3\font minus
  \fontdimen4\font\relax}
\providecommand{\BIBforeignlanguage}[2]{{%
\expandafter\ifx\csname l@#1\endcsname\relax
\typeout{** WARNING: IEEEtran.bst: No hyphenation pattern has been}%
\typeout{** loaded for the language `#1'. Using the pattern for}%
\typeout{** the default language instead.}%
\else
\language=\csname l@#1\endcsname
\fi
#2}}
\providecommand{\BIBdecl}{\relax}
\BIBdecl

\bibitem{Glendinning2017}
I.~Glendinning, T.~Foltynek, and J.~Rybi{\v{c}}ka, \emph{{Plagiarism Across
  Europe and Beyond 2017: Conference Proceedings, May 24-26, 2017, Brno, Czech
  Republic}}.\hskip 1em plus 0.5em minus 0.4em\relax Mendel University, 2017.

\bibitem{Cosma2006}
G.~Cosma and M.~Joy, ``{Source-code plagiarism: A UK academic perspective},''
  \emph{Research Report RR-422. University of Warwick}, 2006.

\bibitem{Culwin2001}
F.~Culwin, A.~MacLeod, and T.~Lancaster, ``{Source code plagiarism in UK HE
  computing schools},'' in \emph{Proc. of. 2nd Annual LTSN-ICS Conference,
  London}, London, 2001.

\bibitem{Bamford2005}
J.~Bamford, K.~Sergiou \emph{et~al.}, ``{International students and plagiarism:
  An analysis of the reasons for plagiarism among international foundation
  students},'' \emph{Investigations in University Teaching and Learning},
  vol.~2, no.~2, pp. 17--22, 2005.

\bibitem{Doss2016}
D.~A. Doss, R.~Henley, B.~Gokaraju, D.~McElreath, H.~Lackey, Q.~Hong, and
  L.~Miller, ``Assessing domestic vs. international student perceptions and
  attitudes of plagiarism,'' \emph{Journal of International Students}, vol.~6,
  no.~2, pp. 542--565, 2016.

\bibitem{Gow2014}
S.~Gow, ``{A cultural bridge for academic integrity? Mainland Chinese
  master’s graduates of UK institutions returning to China},''
  \emph{International Journal for Educational Integrity}, vol.~10, no.~1, pp.
  70--83, 2014.

\bibitem{Shei2005}
C.~Shei, ``{Plagiarism, Chinese learners and Western convention},''
  \emph{Taiwan Journal of TESOL}, vol.~2, no.~1, pp. 97--113, 2005.

\bibitem{Boniecki2009}
M.~Boniecki, Z.~Gniazdowski, and T.~Nowakowski, ``Wprowadzenie w
  rozwi{\c{a}}zywanie konkursowych zada{\'n} programistycznych,'' \emph{Zeszyty
  Naukowe WWSI}, vol.~3, no.~3, pp. 9--22, 2009.

\bibitem{Ali2011}
A.~M. E.~T. Ali, H.~M.~D. Abdulla, and V.~Sn{\'a}{\v{s}}el, ``{Overview and
  Comparison of Plagiarism Detection Tools},'' \emph{Dateso 2011}, pp.
  161--172, 2011.

\bibitem{Hage2010}
J.~Hage, P.~Rademaker, and N.~van Vugt, ``A comparison of plagiarism detection
  tools,'' \emph{Utrecht University. Utrecht, The Netherlands}, vol.~28, 2010.

\bibitem{Heres2017}
\BIBentryALTinterwordspacing
D.~Heres and J.~Hage, ``{A Quantitative Comparison of Program Plagiarism
  Detection Tools},'' in \emph{Proceedings of the 6th Computer Science
  Education Research Conference}, ser. CSERC '17.\hskip 1em plus 0.5em minus
  0.4em\relax New York, NY, USA: ACM, 2017, pp. 73--82. [Online]. Available:
  \url{http://doi.acm.org/10.1145/3162087.3162101}
\BIBentrySTDinterwordspacing

\bibitem{Lukashenko2007}
R.~Lukashenko, V.~Graudina, and J.~Grundspenkis, ``Computer-based plagiarism
  detection methods and tools: an overview,'' in \emph{Proc. of the 2007
  International Conference on Computer Systems and Technologies}.\hskip 1em
  plus 0.5em minus 0.4em\relax Citeseer, 2007.

\bibitem{Martins2014}
V.~T. Martins, D.~Fonte, P.~R. Henriques, and D.~da~Cruz, ``{Plagiarism
  detection: A tool survey and comparison},'' in \emph{3rd Symposium on
  Languages, Applications and Technologies}.\hskip 1em plus 0.5em minus
  0.4em\relax Schloss Dagstuhl-Leibniz-Zentrum fuer Informatik, 2014.

\bibitem{Dasgupta2008}
S.~Dasgupta, C.~H. Papadimitriou, and U.~V. Vazirani, \emph{Algorithms}.\hskip
  1em plus 0.5em minus 0.4em\relax McGraw-Hill Higher Education, 2008.

\bibitem{Deza2009}
M.~M. Deza and E.~Deza, \emph{Encyclopedia of distances}.\hskip 1em plus 0.5em
  minus 0.4em\relax Springer, 2009.

\bibitem{Ross1992}
K.~A. Ross and C.~R.~B. Wright, \emph{Discrete mathematics}.\hskip 1em plus
  0.5em minus 0.4em\relax Prentice Hall, 1992.

\bibitem{Peters2012}
J.~F. Peters and P.~Wasilewski, ``Tolerance spaces: Origins, theoretical
  aspects and applications,'' \emph{Information Sciences}, vol. 195, pp.
  211--225, 2012.

\bibitem{Mason2018}
T.~Mason, A.~Gavrilovska, and D.~A. Joyner, ``{Collaboration Versus Cheating.
  Reducing Code Plagiarism in an Online MS Computer Science Program},''
  \emph{arXiv preprint arXiv:1812.00276}, 2018.

\end{thebibliography}
\bibliographystyle{IEEEtran}
\end{document}